\documentclass[preprint, a4paper, times, modern]{aastex62}
\usepackage{amsmath}
\usepackage{graphicx}
\usepackage{natbib}

\newcommand{\kms}{~km~s$^{-1}$}

\begin{document}
\title{On the sensitivity of heliosphere models to the uncertainty of the low-energy charge exchange cross section}

\correspondingauthor{M. Bzowski}
\email{bzowski@cbk.waw.pl}

\author[0000-0003-3957-2359]{M. Bzowski}
\affiliation{Space Research Centre, Polish Academy of Sciences (CBK PAN),\\
Bartycka 18A, 00-716 Warsaw, Poland}

\author[0000-0001-7867-3633]{J. Heerikhuisen}
\affiliation{University of Waikato,\\
Hamilton, New Zealand}

\begin{abstract}

Models play an important role in our understanding of the global
structure of the solar wind and its interaction with the interstellar
medium. A critical ingredient in many types of models are the
charge-exchange collisions between ions and neutrals. Some ambiguity
exists in the charge-exchange cross-section for protons and hydrogen
atoms, depending on which experimental data is used. The differences
are greatest at low energies, and for the plasma-neutral interaction
in the outer heliosheath may exceed 50\%. In this paper we assess a
number of existing data sets and formulae for proton-hydrogen
charge-exchange. We use a global simulation of the heliosphere to
quantify the differences between the currently favored cross-section,
and a formulation we suggest that more closely matches the majority of
available data. We find that in order to make the resulting two
heliospheres the same size, the interstellar proton and hydrogen
densities need to be adjusted by 10 to 15\%, which provides a way to link the uncertainty in the   cross-section to the uncertainty in the parameters of the pristine interstellar plasma.
  
\end{abstract}

\section{Introduction}
\label{sec:intro}
The heliosphere is created due to the interaction between the solar wind plasma and both the charged and the neutral components of the local interstellar medium (LISM). At the spatial scale of the heliosphere, the neutral LISM component is weakly collisional. Therefore, modeling of the interaction responsible for the creation of the heliosphere should be done using an MHD model for the plasma-plasma interaction and a kinetic treatment of the neutral-plasma coupling, with the collision-energy-dependent cross section taken into account \citep{baranov_malama:93, izmodenov_etal:05a, heerikhuisen_etal:06a, heerikhuisen_etal:08a}. The resulting exchange of momentum and energy between the plasma and neutral components in the outer heliosheath OHS -- the region of interstellar plasma that is affected by the presence of the heliosphere -- due to charge exchange collisions have been shown to be important processes affecting the size of the heliosphere and physical state (density, temperature, flow speed and direction) in the inner and outer heliosheath and the distribution function of interstellar neutral H penetrating inside the termination shock \citep{baranov_etal:98a, heerikhuisen_etal:16a}. The rate of charge exchange reactions depends, among others, on the magnitude of the charge exchange cross section and on its variation with collision speed. 

The charge exchange cross section formula used in heliospheric research during the past decade has been widely adopted after \citet{lindsay_stebbings:05a} (LS05), who derived it based on a compilation of measurements for the range of collision energies from $\sim 0.005$ to $\sim 200$~keV. In particular, the LS05 formula has been used in well known simulation models of the heliosphere by e.g. \citet{heerikhuisen_pogorelov:10a, izmodenov_alexashov:15a, czechowski_grygorczuk:17a, opher_etal:15a}. However, measurements of this cross section are challenging for low energies, and results from different experiments are sometimes discrepant beyond the uncertainty ranges. Therefore, several alternative formulae have been used in the past, depending on the choice of experimental data. The most widely adopted among these formulae were those from \cite{fite_etal:62}, \cite{maher_tinsley:77}, and \citet{barnett_etal:90} (Ba90). 

\citet{baranov_etal:98a} demonstrated the sensitivity of results of kinetic models of the heliosphere to the adopted dependence of the charge exchange cross section on collision speed. They compared models calculated with the cross section from \cite{maher_tinsley:77} with that from \citet{fite_etal:62}. The magnitudes of these two cross sections differ approximately 20\% in the entire range of collision speeds (from a few \kms to 500\kms). \citet{baranov_etal:98a} found that the densities of ISN H both in the OHS and inside the heliosphere differ by $\sim 15$\% but otherwise the shape of the heliosphere and the locations of the heliopause and the termination shock vary very little (about 1\%).

Here, we show that the charge-exchange cross section obtained from the formula from LS05 systematically differs from certain important measurements of this quantity in the energy range characteristic for the outer heliosheath (OHS), while it agrees with others. Since we are unable to determine which of the data sets are correct, and hence which analytical approximations of the charge exchange cross section better represent the reality, in this paper we seek to understand the effect of the aforementioned uncertainty in the low-energy cross section for charge exchange between H atoms and protons on the results of modeling of the heliosphere, and in particular the outer heliosheath in the upwind hemisphere. We use the well-established Huntsville MHD model of the heliosphere with kinetic treatment of the neutral gas-plasma interaction \citep{pogorelov_etal:09e, heerikhuisen_pogorelov:10a}. We compare results of the model run with identical parameters and with either the LS05 charge exchange formula or a formula that we fit here to the low-energy charge exchange  measurements recommended by Ba90. We show the differences in the locations of the termination shock and heliopause, as well as the outer heliosheath. We discuss differences between the plasma flow parameters in the OHS and the parameters of the modified hydrogen population penetrating inside the termination shock. In addition, we show what changes to the LISM hydrogen and proton densities are needed with the Ba90 cross-section in order to obtain a similarly constrained heliosphere as is obtained using the LS05 cross-section.

\section{Measurements and models of the charge exchange cross section}
\label{sec:cxCrosssect}

The collision energy range relevant for global modeling of the heliosphere is from $\sim 1$~eV to $\sim 6$~keV (10--1000~km~s$^{-1}$). This is because collision speeds in the outer heliosheath vary from about 10\kms\, for a cold H atom co-moving with the flow of plasma at 7500~K to $\sim 70$\kms\, for a H atom running at 30\kms\, across plasma at $3\times 10^4$~K. On the other side of the interest range, in the supersonic solar wind inside the termination shock, the fast solar wind expands in the polar regions expands at 750--1000\kms\,\citep{phillips_etal:95c}, which are characteristic collision speeds between solar wind protons and interstellar H atoms inside the termination shock in the polar regions during solar minimum conditions. This is the speed range that a formula for charge exchange cross section must be valid for in order to provide accurate models for plasma-neutral interactions in the heliospheric interface.

\begin{figure*}
\epsscale{1.1}
\plottwo{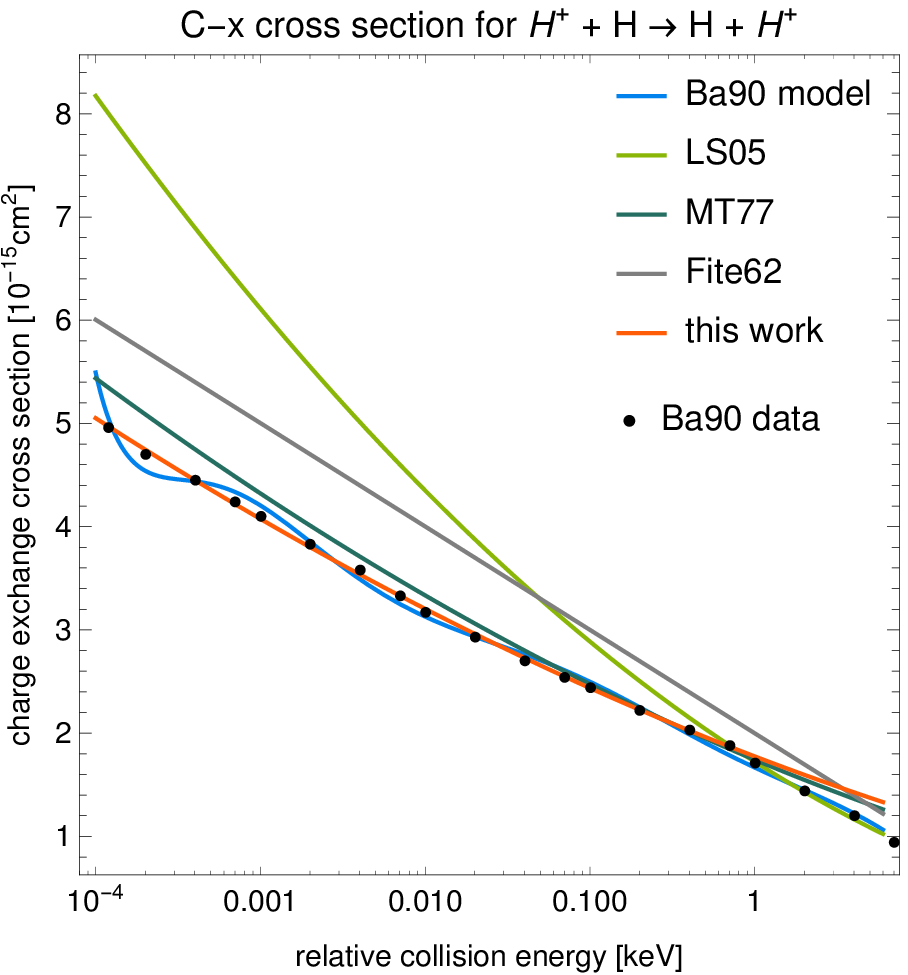}{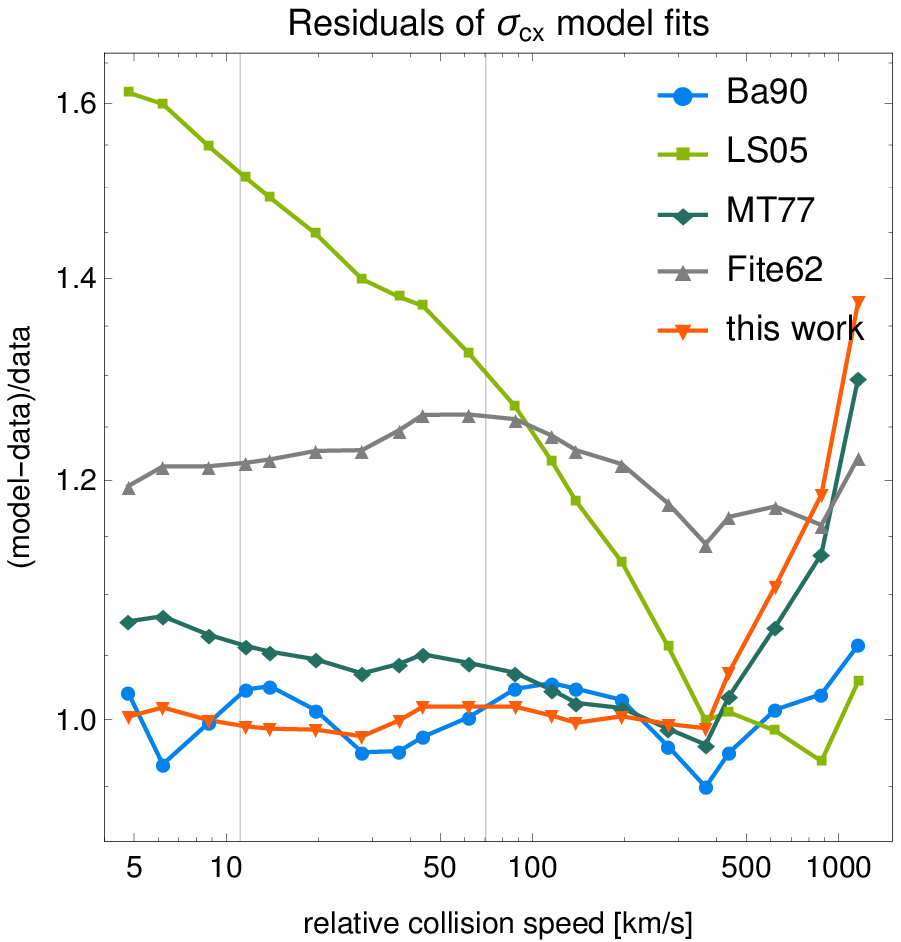}
\caption{Comparison of measurements of the H + H$^+$ charge exchange cross section as a function of collision energy, recommended by \citet{barnett_etal:90} (black dots) with approximation formulae from \citep[][ Ba90]{barnett_etal:90}, \citep[][ LS05]{lindsay_stebbings:05a}, \citep[][ MT77]{maher_tinsley:77}, \citep[][ Fite62]{fite_etal:62}, and this work  corresponding to Equation~\ref{eq:CXFormula} (left panel) and the residuals of these formulae as a function of collision speeds (right panel). The energy scale in the left panel precisely corresponds to the collision speed scale in the right panel.}
\label{fig:cxCompar}
\end{figure*}

Charge exchange cross section measurement data are compiled and approximation formulae are suggested by Ba90 and LS05. Ba90 provides the recommended data in a numerical form, while LS05 only plots the data sets they used, but the scale of the figure in this paper does not facilitate extracting the values with a sufficient accuracy. The measurements recommended by Ba90 are in agreement with recent theoretical models \citep[e.g.,][]{kadyrov_etal:06a}, and the cross section values obtained from Ba90 and LS05 agree for collision speeds above $\sim 300$\kms. However, the magnitude of differences between the LS05 model values and the measurements recommended by Ba90 increases with a decreasing energy from $\sim 15$\% at $200$\kms\, ($\sim 200$~eV) to $\sim 50$\% at 10\kms\, ($\sim 0.5$~eV; see the right panel in Figure~\ref{fig:cxCompar}). 

In the energy range relevant for the OHS, the measurements used by LS05 must have been adopted from \citet{belyaev_etal:67a}, while Ba90 recommend values consistent with \citet{newman_etal:82}. These two measurement papers used a similar experimental technique but obtained different results for the cross sections in the lowest energy range (see Figure~2 in \citet{belyaev_etal:67a} and Figure~5 in \citet{newman_etal:82}). In both of these papers, the measurements are compared with theoretical models. The models used in both papers agree that the cross section should linearly increase with the logarithm of decreasing energy. However, while the measurements and the model in \citet{newman_etal:82} fit well to each other, the data obtained by \citet{belyaev_etal:67a} start to deviate from the model quite abruptly as the interaction energy decreases below $\sim 100$~eV. Another comparison of a newer model with the measurements is provided by \citet{kadyrov_etal:06a}, who demonstrates that their model fits the measurement from \citet{newman_etal:82} down to 1~eV. The deviation of the values returned by LS05 from the measurements by \citet{newman_etal:82} starts at 300\kms\, ($\sim 470$~eV), and at 10\kms\, it attains 50\% of the measured value. 

For low collision speeds (up to $\sim 400$\kms), \citet{maher_tinsley:77} suggested (and LS05 reiterated) that a good functional approximation for the dependence of the charge exchange cross section on collision energy is 
\begin{equation}
\label{eq:MTCXFormula}
\sigma_{cx}(E) = \left(a + b \ln E\right)^2
\end{equation}
This formula was also adopted by \citet{izmodenov_alexashov:15a}, who fitted the coefficients of this equation to the LS05 formula and used it in the Moscow Monte Carlo model of the heliosphere. We took the recommended data from Ba90 included within the speed range from 4.79 to 368\kms and fitted the parameters from Equation~\ref{eq:MTCXFormula} to obtain the following charge-exchange cross-section formula, where $E$ is collision energy in eV and $\sigma_{cx}$ is the cross section in cm$^2$.
\begin{equation}
\label{eq:CXFormula}
\sigma_{cx}(E) = \left(6.384\times10^{-8} - 3.14\times10^{-9} \ln E\right)^2
\quad\mbox{for}\quad
10^{-4} < E < 1 \mbox{keV}.
\end{equation}
This cross-section is plotted in the left panel of Figure~\ref{fig:cxCompar}, along with the data and model recommended by Ba90 and model predictions by LS05, \citet{maher_tinsley:77}, and \citet{fite_etal:62} presented as a function of collision energy. Residuals of these model values are presented in the right-hand panel of this figure.  

Clearly, the LS05 model deviates from the data between 30 and 60\% in the energy range characteristic for the charge exchange collisions in the OHS. The model by \citet{fite_etal:62} deviates upward almost   uniformly by $\sim 20$\%. The other presented models agree with the data within this range to $\sim 10$\%. The best agreement between the data and all models occurs close to 400~\kms, i.e., for collision speeds characteristic for slow solar wind. For larger velocities, LS05 and Ba90 agree with  the data very well (within $\sim 10$\%), while MT77 and (\ref{eq:CXFormula}) abruptly deviate. In the simulations shown in the following section, we use Equation~\ref{eq:CXFormula} for energies up to 1 keV, and LS05 at higher energies.
 
Because of the large differences between the LS05 the other models for the low collisions speeds, it can be expected that the coupling between the neutral gas and the plasma in the OHS in the simulations using these different cross section models will be different, and therefore results of the global heliospheric models will be different. Differences are likely to appear in the simulated locations of the heliopause and the termination shock, in the flow and temperature of the plasma in the OHS, as well as in the production of the secondary components of interstellar neutral atoms, both hydrogen \citep[e.g,][]{izmodenov:00}, helium \citep{bzowski_etal:17a} and other species. Potentially, an unaccounted dependence of the model on the charge exchange cross section may bias important physical parameters of the heliosphere and the LISM derived using the heliosphere models. Therefore, we find it compelling to investigate the uncertainties of some key quantities obtained from heliospheric models due to the uncertainty in the low-energy charge exchange cross section. Specifically, by using simulations that employ either the LS05 or Ba90 cross-section, we are able to quantify the differences in LISM conditions needed to reproduce key observable like the distance to the heliopause and the density of neutral hydrogen in the solar wind.

\section{Simulation model of the heliosphere}
\label{sec:helioModel}
We used the Huntsville model of the heliosphere, also known as MS-FLUKSS \citep{pogorelov_etal:09e}. For the simulation used here we employed a spherical grid with kinetic neutrals, and assumed a kappa-distribution for protons in the inner heliosheath with a kappa index of 1.63 \citep[e.g.][]{heerikhuisen_etal:08a}. Since we are particularly interested in the charge-exchange process, our code computed the charge-exchange rate for a given H-atom by keeping the cross-section inside the collision integral over the local proton distribution \citep{Heerikhuisen_etal:15a}. Most other heliospheric models utilize some form of average interaction speed \citep[e.g.][]{pauls_etal:95,mcnutt_etal:98}, which work reasonably well when the proton temperature is low, and when protons are approximately Maxwellian \citep{destefano_heerikhuisen:17a}.

For the solar wind in the simulations, we assume that the entire volume inside the termination shock is filled with the slow wind with a density of 6.55~nuc~cm$^{-3}$ and a flow speed of 387~\kms at 1~au , which gives a dynamic pressure similar to what was observed during the 2010 to 2015 period in the ecliptic plane \citep{mccomas_etal:18a}. The simulated solar wind also contains a Parker spiral magnetic field with the strength of 37.5~ $\mu$G at 1~au, though to prevent an unphysical flat current sheet from distorting the heliopause we assume the same polarity in both northern and southern hemispheres.

At the LISM side of the heliosphere, we adopt the magnetic field vector in the unperturbed LISM from a recent analysis of the IBEX ribbon ENAs by \citet{zirnstein_etal:16b} with a field strength of 2.93~$\mu$G.The LISM temperature and the vector of the Sun's velocity (i.e., the inflow velocity of interstellar matter into the heliosphere) were taken from direct-sampling measurements of the primary component of interstellar He observed by Ulysses \citep{bzowski_etal:14a} and IBEX \citep{bzowski_etal:15a, mccomas_etal:15b}. We adopted $T = 7500$ K and $V = 25.4$~\kms in the LISM. Given this configuration, we adjusted the interstellar plasma and neutral densities until we obtained a heliosphere that satisfies the heliopause location as observed at 121~au by the \emph{Voyager 1} spacecraft \citep{gurnett_etal:13a}, and a density of neutral hydrogen at the heliospheric termination shock of $\sim 0.087$ cm$^{-3}$ as implied by analysis of pickup ion observations on Ulysses \citep{bzowski_etal:08a} and the solar wind slowdown due to the pickup of interstellar neutral hydrogen atoms ionized inside the termination shock \citep{richardson_etal:08a}. 

We ran three different simulations of the solar wind interacting with the heliosphere using the model and boundary conditions just described. The first model utilizes the LS05 cross-section, which we will refer to as $\sigma_{\rm LS05}$. To satisfy the requirements on the heliopause distance and the hydrogen density in the solar wind, we require LISM densities $n_p = 0.064$ cm$^{-3}$ and $n_H = 0.132$ cm$^{-3}$. These densities, combined with the LISM and solar wind properties described above, comprise what we call the BC$_{\rm LS05}$ boundary conditions for this model heliosphere. This heliosphere represents the baseline of our numerical investigation.

Next, we repeated the simulation, with the only change being a switch from the LS05 cross-section to the Ba90 cross-section, $\sigma_{\rm  Ba90}$, which uses Equation \ref{eq:CXFormula} below 1 keV and LS05 at higher energies, thereby keeping the BC$_{\rm LS05}$ boundary conditions. The resulting heliosphere is different from the baseline since the charge-exchange rates, especially in the OHS, are different. As a result, the requirements on the heliopause location and hydrogen density at the termination shock are no longer satisfied. 

Finally, we present a model heliosphere that was obtained after modifying the LISM boundary conditions (through trial and error) such that we obtain approximately the same heliopause distance and hydrogen density inside the heliosphere as the baseline. Using the $\sigma_{\rm Ba90}$ cross-section this requires LISM densities $n_p = 0.075$ cm$^{-3}$ and $n_H = 0.122$~cm$^{-3}$, which, combined with the other boundary conditions that were left unchanged, comprise what we refer to as the BC$_{\rm Ba90}$ boundary conditions.
 
\section{Results}
\label{sec:results}

\begin{figure*}
\epsscale{1.2}
\plotone{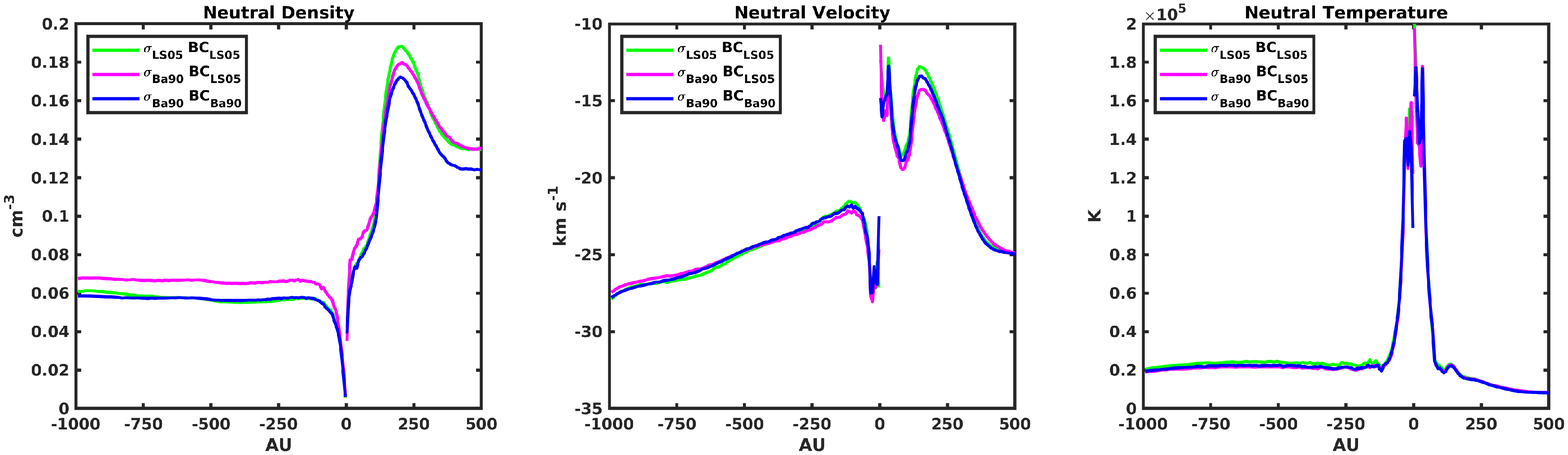}
\caption{Bulk neutral hydrogen properties along the radial line from the Sun that is parallel to the Sun's motion through the LISM. Here we plot the number density (left), velocity component along the line (middle), and temperature (right), for each of the three cases considered. These quantities were computed by taking moments of the particle distribution function in the code.}
\label{fig:neutral_lines}
\end{figure*}

Figure \ref{fig:neutral_lines} shows the bulk properties of neutral hydrogen along  the direction of motion of the Sun through the LISM. Switching to the Ba90 cross-section causes the peak neutral density in the hydrogen wall (OHS) to be about 5\% lower, while the neutral density inside the heliosphere is about 15\% higher than the baseline case. This is due to the lower rate of charge-exchange obtained in the OHS with the Ba90 cross-section, which allows more interstellar neutrals to pass unimpeded into the heliosphere.

\begin{figure*}
\epsscale{1.2}
\plotone{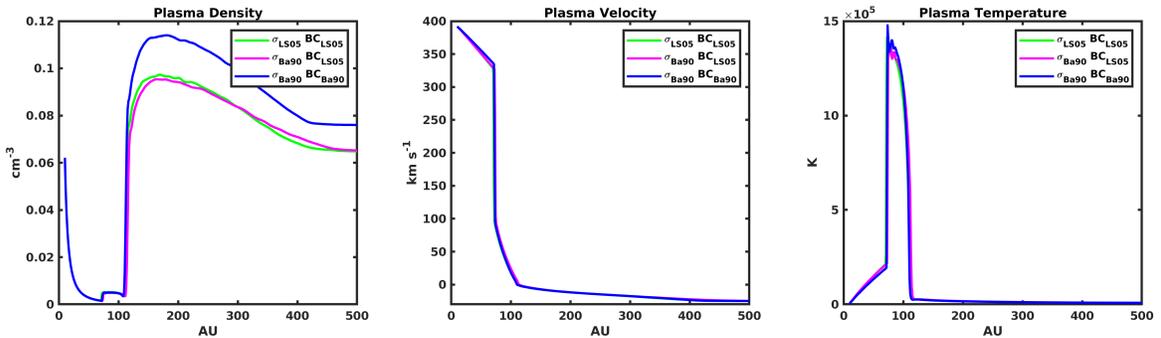}
\caption{Bulk proton properties along the radial line from the Sun that is parallel to the Sun's motion through the LISM, obtained from the MHD solver used in the code. Here we plot the number density (left), the velocity component parallel to the line (middle), and the plasma temperature (right), for each of the three cases considered.}
\label{fig:plasma_lines}
\end{figure*}

While the charge-exchange rates differ significantly in the OHS, where the temperatures and the flow speeds are low, inside the termination shock the two cross-sections are more similar. As Figure \ref{fig:cxCompar} shows, however, the Ba90 cross-section is still smaller, so despite having more neutrals inside the solar wind, the slow-down and heating of the supersonic solar wind due to charge-exchange end up being about the same.
The lower rate of charge-exchange in the OHS also results in less slowing of the incoming neutrals inside the hydrogen wall (middle plot of Figure \ref{fig:neutral_lines}).
Interestingly, the case with the Ba90 cross-section has a slightly larger OHS as compared to the baseline, with the plasma density increasing above the LISM value around 450 au, which is about 30-50 au sooner than the LS05 cross-section case (left plot in Figure \ref{fig:plasma_lines}). This effect is due to the slightly smaller Ba90 cross-section that gives the neutral solar wind -- neutrals born through charge-exchange in the supersonic solar wind -- a slightly larger mean free path in the OHS as these particle escape the heliosphere.

\begin{figure*}
\epsscale{1.2}
\plotone{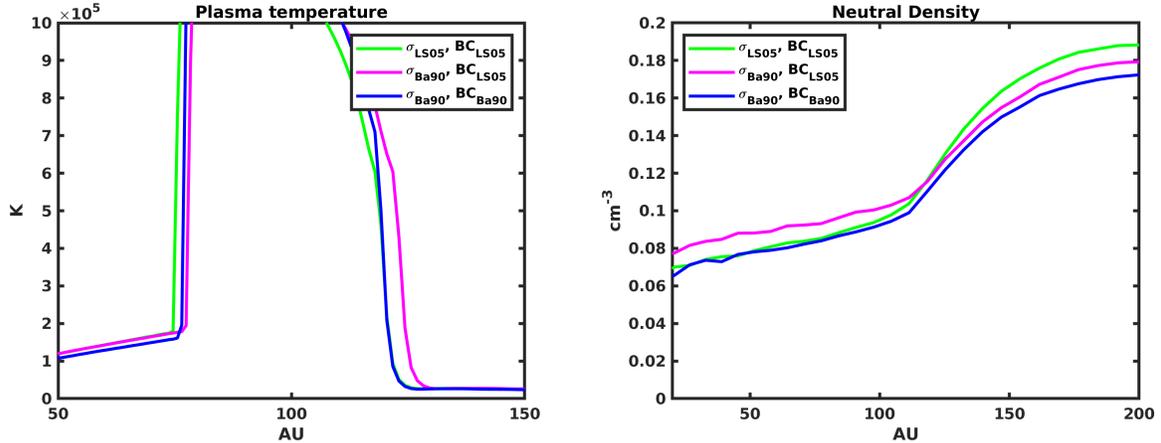}
\caption{Plots illustrating {\bf{the effects of }} the boundary condition on the size of the heliosphere (left) and neutral hydrogen density inside the heliosphere (right). Left we have plotted the plasma temperature along the radial line from the Sun through the \emph{Voyager 1} location, which shows that both the $\sigma_{\rm LS05}$ with BC$_{\rm LS05}$ and the $\sigma_{\rm Ba90}$ with BC$_{\rm Ba90}$ have the same heliopause location at approximately 120 au. The right plot is along the radial line from the Sun that is parallel to the Sun's motion through the LISM, and shows that both the $\sigma_{\rm LS05}$ with BC$_{\rm LS05}$ and the $\sigma_{\rm Ba90}$ with BC$_{\rm Ba90}$ have approximately the same neutral hydrogen density in the supersonic solar wind.}
\label{fig:zoomed_lines}
\end{figure*}

Overall, the effect of switching to the Ba90 cross-section while keeping the BC$_{\rm LS05}$ boundary conditions is to reduce the momentum exchange from neutrals to the plasma in the OHS, which in turn allows the heliosphere to expand slightly. We find that the termination shock and heliopause move out about 2 au and 3 au, respectively, in the direction of the nose of the heliosphere (see also Figure \ref{fig:zoomed_lines}).

We then switch to the case with the $\sigma_{\rm Ba90}$ and BC$_{\rm Ba90}$ boundary conditions. Figure \ref{fig:zoomed_lines} shows that in this case the heliopause moves back to the same location as the baseline, and the hydrogen density in the solar wind is also approximately consistent with the baseline. So while the inside of the heliosphere is similar, we have had to significantly change the LISM conditions. Most striking is the region of plasma just outside the heliosphere whose profile is about 17\% higher than the baseline case. Interestingly, this is about the same fraction that the neutral density increased by inside the heliosphere under the $\sigma_{\rm Ba90}$ with BC$_{\rm LS05}$ conditions. But since this is not a linear feedback system, we also had to reduce the LISM hydrogen density by about 8\%.

A comparison of the influence of the magnitude of charge exchange cross section was made by \citet{baranov_etal:98a}. Our conclusions are similar to theirs concerning the location of the heliopause and the termination shock: for all other parameters of the model unchanged, the termination shock and the heliopause change locations just by 1 -- 3\%. However, unlike \citet{baranov_etal:98a}, we found that adoption of $\sigma_{\rm Ba90}$ instead of $\sigma_{\rm LS05}$ results in modifying the relation between the H density in the OHS and inside the termination shock. We believe that this is because the relations between the alternative cross section used by \cite{baranov_etal:98a} and by us were different: in the former case, the ratio of the {\bf {two alternative cross sections}} was very weakly dependent on the collision speed, in the latter case, the cross section ratio was increasing from 1 for the typical solar wind speed of $\sim 440$~\kms{} to $\sim 1.5$ at the lower end of the collision speed range in the OHS.

\section{Summary and conclusions}
\label{sec:conclusions}

We investigated a number of past works on the hydrogen-proton cross-section for charge-exchange collisions and found that the commonly used LS05 cross-section differs significantly from the others for energies below about 1 keV. To determine the impact this might have on how we interpret various ion-neutral reactions in the heliospheric interface, we ran simulations of the global heliosphere for three different cases: 1) a baseline case using the LS05 cross-section with boundary conditions that result in a heliosphere that meets observation-based constraints on its size and properties; 2) the same boundary conditions as (1), but with the Ba90 cross-section; 3) a run with the Ba90 cross-section, but where we modified the LISM densities such that the observational constraints from (1) are also satisfied.

Compared to the baseline case, switching to the Ba90 cross-section reduces the charge-exchange rate in the outer heliosheath, which lets more neutral hydrogen into the heliosphere. Reduced charge-exchange leads to less momentum transfer onto the outer heliosheath plasma by interstellar neutrals, which allows the heliopause to move out by $\sim 3$ au. In the third case we increase the interstellar plasma density to ensure that the resulting increase in charge-exchange pushes the heliopause back to the baseline location. The ratio of the two cross-sections is not constant with energy, and also has a small but non-trivial effect at energies of a few hundred eV. This requires to also reduce the LISM hydrogen density in order to match the baseline conditions. Even with the same amount of hydrogen inside the heliosphere, the third case differs slightly from the baseline since the charge-exchange rate in the slow solar wind is lower with the Ba90 cross-section, which pushes the heliospheric termination shock out by ~1 au.

Overall we have shown that the form of the charge-exchange cross-section significantly affects the interpretation of models used to understand the structure of the heliosphere. Switching between the LS05 and Ba90 cross-sections has a similar impact on the heliosphere as changing the LISM densities by $\sim 10-15$\%. As a result, we advocate that the community apply the LS05 cross-section only above 1~keV, while below that we suggest using Equation~\ref{eq:CXFormula}.

\acknowledgments
{\emph{Acknowledgments}}. This study was supported by Polish National Science Center grant 2015-18-M-ST9-00036.

\bibliographystyle{aasjournal}
\bibliography{iplbib}

\end{document}